\documentclass{sf2a-conf2015}
\usepackage{graphicx}
\usepackage{hyperref}
\usepackage[]{natbib}  
\usepackage{epstopdf}

\def\BibTeX{{\rm B\kern-.05em{\sc i\kern-.025em b}\kern-.08em
		T\kern-.1667em\lower.7ex\hbox{E}\kern-.125emX}}
\bibpunct{(}{)}{;}{a}{}{,}  

\begin{document}
	
	\TitreGlobal{proceeding of the SF2A 2015. 7 pages, 2 Fig}
	
	
	\title{Mapping the inner stellar halo of the Milky Way from 2MASS and SDSS-III/APOGEE survey}
	
	\runningtitle{Stellar halo of the Milky Way}
	
	\author{J. G. Fern\'andez-Trincado}\address{Institut Utinam, CNRS UMR 6213, Universit\'e de Franche-Comt\'e, OSU THETA Franche-Comt\'e-Bourgogne, Observatoire de Besan\c{c}on, BP 1615, 25010 Besan\c{c}on Cedex, France.}
	
	\author{A. C Robin$^1$}
	\author{C. Reyl\'e$^1$}

	
	

	\setcounter{page}{237}


\maketitle


\begin{abstract}
The Besan\c{c}on Galaxy model was used to compare the infrared colour distribution of synthetic stars with those from 2MASS observations taking the selection function of the data into account, in order to study the shape of the stellar halo of the Milky Way, with complemetary spectroscopic data from SDSS-III/APOGEE survey. Furthermore, we compared  the generated mock metallicity distribution of the Besan\c{c}on Galaxy model, to the  intrinsic metallicity distribution with reliable stellar parameters from the APOGEE Stellar Parameters and Chemical Abundances Pipeline (ASPCAP). The comparison was carried accross a large volume of the inner part of the Galaxy, revealing that a metal-poor population, [M/H]$<-1.2$ dex, could fill an extended component of the inner galactic halo. With this data set, we are able to model a more realistic mass density distribution of the stellar halo component of the Milky Way, assuming a six-parameters double power-law model, and reconstruct the behaviour of the rotation curve in the inner part of the Galaxy.
\end{abstract}


\section{Introduction}

	  To perform detailed studies on the kinematics of stars in the Milky Way and its components, as well as to interprete the upcoming six-dimensional phase-space data set produced by the Gaia space mission, a more elaborated description of the Galactic potential of the Milky Way is required. With this purpose, and taking advantage of the well described density profiles for each component of the Besan\c{c}on Galaxy model \citep{Robin2003, Robin2012, Robin2014}, an axisymmetric three-dimensional model for the gravitational field for the inner part of the Galaxy (triaxial bar, stellar halo, central mass), is currently modeled \citep[e.g.,][2015 in prep.]{Fernandez-Trincado2014}, and will be tested using the available spectroscopic surveys as SDSS-III/APOGEE \citep{Alam2015}. The aim of our study is constraint the formation scenarios of the Milky Way central regions, and determine whether the Galactic bulge was predominantely formed by mergers according to Cold Dark Matter (CDM) theory \citep[e.g.,][]{Abadi2003}, or from disk instabilities \citep[e.g.,][]{Athanassoula2005}, as suggested by its boxy/peanut shape, or if both processes could have affected the inner regions of the Galaxy.\\
  
  In this paper, we compare the metallicity distribution function of halo (metal-poor) stars predicted by the Besan\c{c}on Galaxy model with SDSS-III/APOGEE spectroscopic data to infer a constraint on the inner halo density laws. 
  

\section{The Milky Way stellar halo}
\label{shapes}

In this section, a brief description of the stellar halo model is outlined.\\

\citet{Robin2014} has recently suggested that a
transition between the Galactic disk and halo of the Milky Way could be smooth enough in the colour-magnitude diagrams, with a halo component dominant at fainter magnitudes, which contribution depends its geometrical shape. In this sense, we attempt to model the contribution of the stellar halo of the Milky Way using a non-spherical (flattened) density double power-law model \citep{Zhao1997} with six free parameters ($\alpha$, $\beta$, $\gamma$, $r_{core}$, $\rho_{\odot}$, $q$),

\begin{equation}
\label{generalfunction1}
\rho (r) = A^{*}  \left(r/r_{core}\right)^{-\gamma}\left[ 1 + \left(r/r_{core}\right) ^{\alpha} \right]^{( \gamma - \beta )/\alpha}
\end{equation}

\begin{equation}
\label{generalfunction2}
 A^{*} =  \rho_{\odot} \left( r_{\odot} / r_{core} \right)^{\gamma} \left[ 1 + \left( r_{\odot} / r_{core} \right)^{\alpha} \right]^{(\beta - \gamma)/\alpha }
\end{equation}\\

\noindent
where $r^2 = X^2 + (Y/p)^2 + (Z/q)^2$ is an axisymmetric radius, and $(X, Y, Z)$ are Galactocentric cartesian coordinates; $r_{core}$ is a scaling radius; $A(\rho_{\odot}, r_{\odot})$  is the normalization, such that, $\rho_{\odot}$ is the local density of the stellar halo in the solar neighborhood ($r_{\odot}=8$ kpc);  $p$ and $q$ are the axis ratios. Axial symmetry ($p=1$) is assumed in this work. The parameters in eq. \ref{generalfunction1} are fitted from 2MASS data \citep{Skrutskie2006}. 
In \citet{Robin2014}, the halo parameters were fitted in the external Galaxy and no constraints on the inner part ($r<4$ kpc) were used. Here we investigate the extrapolation of the density law in the inner Galaxy with different shapes:

\begin{itemize}
	
\item Shape 1 (double power-law): ($\alpha$, $\beta$, $\gamma$, $r_{core}$, $\rho_{\odot}$, $q$) = (1, 3.76, 1, 2180 pc, $0.414\times10^{-4}$ M$_{\odot}/pc^3$, 0.77)

\item Shape 2 (simple power-law): ($\alpha$, $\beta$, $\gamma$, $r_{core}$, $\rho_{\odot}$, $q$) = (1, 2.76, 0, 2180 pc, $0.414\times10^{-4}$ M$_{\odot}/pc^3$, 0.77)

\end{itemize}

\section{Results}

 \subsection{SDSS-III/APOGEE bulge fields: An empirical testbed for the inner Galactic regions}
  
\begin{figure}[ht!]
  	\begin{center}
  		\includegraphics[width=0.7\textwidth]{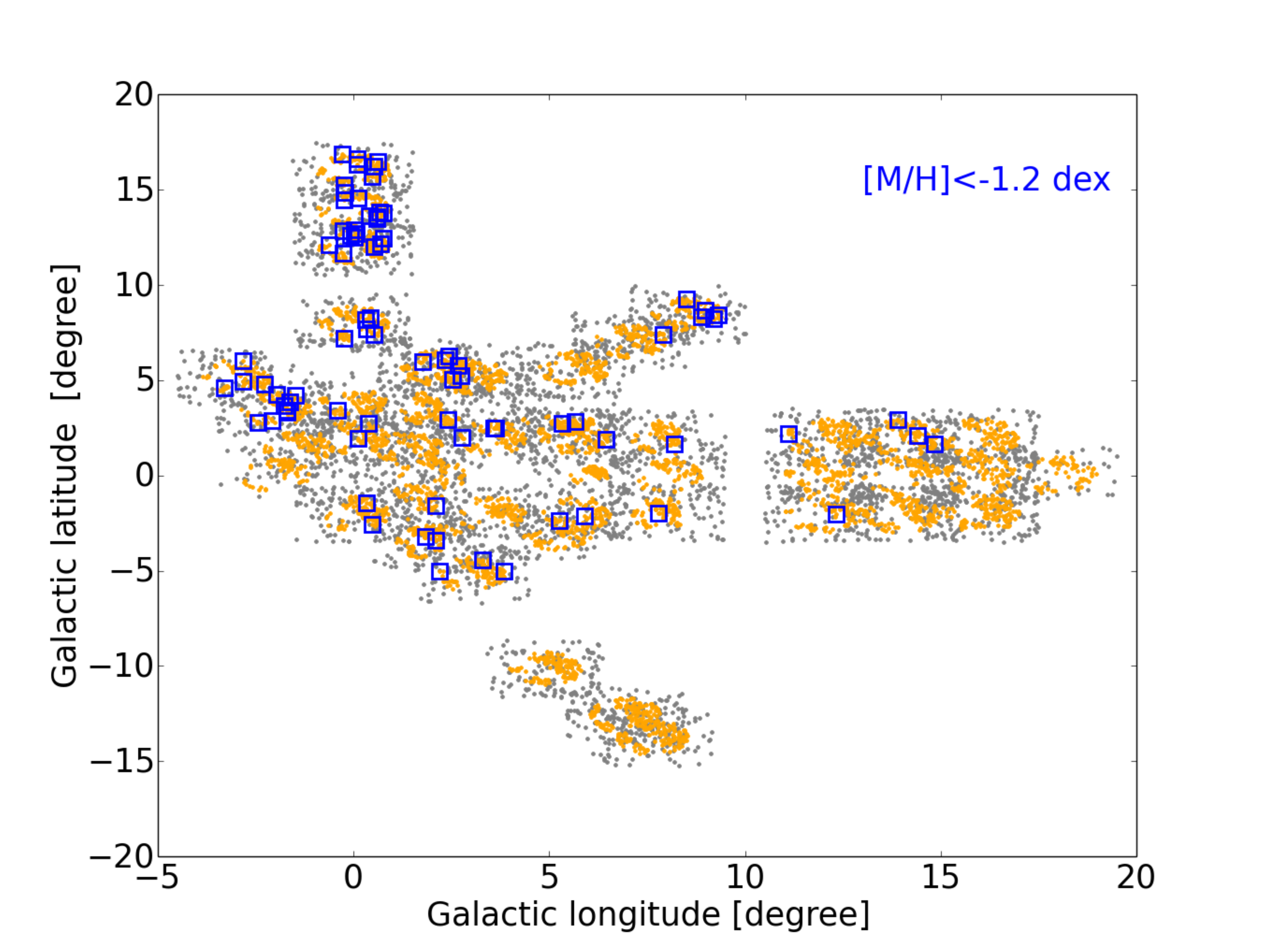}
  	\end{center}
  	\caption{Spatial distribution of 40 SDSS-III/APOGEE fields (orange dots) included in this study, and the Besan\c{c}on Galaxy model simulated APOGEE data (grey dots). The low metallicity stars from SDSS-III/APOGEE are shown as blue squares } 
  	\label{Spatial}
  \end{figure}

\begin{figure}[ht!]
  	\begin{center}
  		\includegraphics[width=0.7\textwidth]{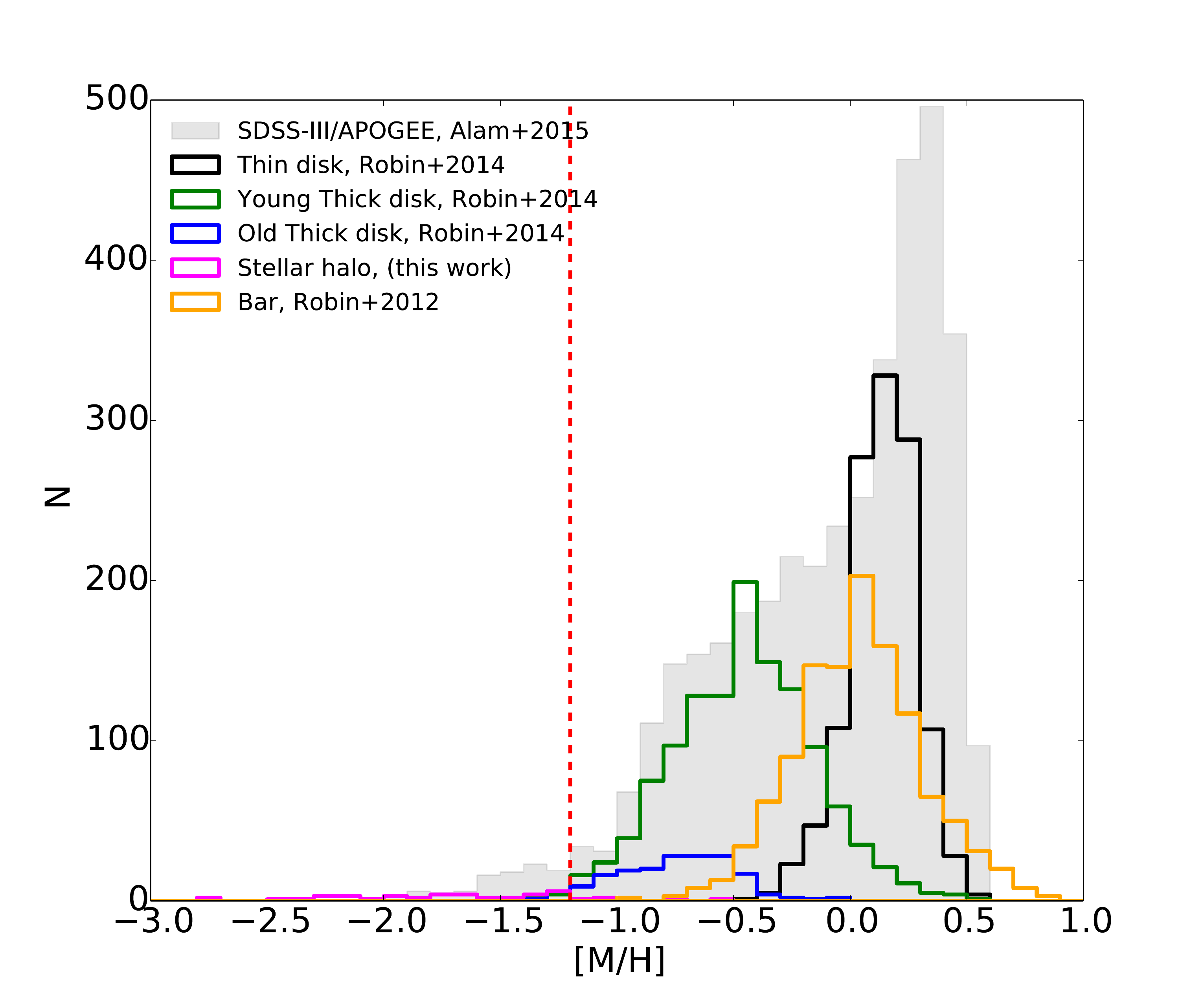}
  	\end{center}
  	\caption{Metallicity distribution functions in 0.1 dex bins for 40 bulge fields.} 
  	\label{MH}
  \end{figure}

  We have selected 40 fields from the SDSS-III/APOGEE database, in order to covers the region defined by -5 deg$ < l < 20$ deg and $ |b| < 20$ deg (see Figure \ref{Spatial}). We select a sample of $\sim$ 4000 stars with high-quality stellar parameters, and control cuts laid out by Garc\'ia  P\'erez A. E. et al. (2015, submitted). The Besan\c{c}on Galaxy model is used with the assumed mass density distribution given in eq. \ref{generalfunction1}, and the parametrized form as in \citet{Robin2014} to produce the expected metallicity distribution function of stellar populations, as shown in Figure \ref{MH}. At low metallicity ([M/H]$<-$1.2 dex) the model expects little contribution from the triaxial bar, thin-disk and more of the Young/Old thick-disk, and our stellar halo. However, the set of parameters with ($\alpha$, $\beta$, $\gamma$, $r_{core}$, $\rho_{\odot}$, $q$) = (1, 3.76, 1, 2180 pc, $0.414\times10^{-4}$ M$_{\odot}/pc^3$, 0.77), does not reproduce the number of stars observed beyond [M/H]$<$-1.2 dex. In the SDSS-III/APOGEE sample it seems that the number of low metallicity stars is much larger than expected from this halo model. To investigate further we show in Figure \ref{Spatial} the distribution in Galactic latitude and longitude of the low metallicity stars. It appears to be not smoothly distributed, but due to variation of extinction in different fields and to the selection function of the APOGEE survey, it is not straightforward to deduce the shape of the stellar halo that could reproduce well the observed distribution. Currently we are fine tuning the parameters of the mass density distribution given in eq. \ref{generalfunction1}, taking into account SDSS-III/APOGEE data of the central region of the Galaxy (Fernandez-Trincado et al. 2015 in prep.).\\
  
  We also expect that a double power-law mass density distribution could be able to explains the observed metallicity distribution function beyond [M/H]$<$ -1.2 dex (Figure \ref{MH}). However, the parametrized form of eq. \ref{generalfunction1} from 2MASS data, does not reproduce the number of stars observed in the tail of the metallicity distribution function.\\
  
 \newpage
  		
\subsection{Building the rotation curve for the inner stellar halo shape: Preliminary results}

In order to reconstruct the potential for the stellar halo, we approximate the triaxial density by a sum of homogeneous spheroidal surfaces, whose densities approximate the mass density distribution in eq. \ref{generalfunction1}, with a step-stair function, according to the adopted stratification method such as that of  \citet{Schmidt1956, Pichardo2004}. The circular velocity $V_{circ}$, for radius ${\bf r_{gal}=r(X,Y,Z=0)}$, is computed as follows: \\

\begin{equation}
\label{generalfunction2}
 V^2_{circ} = {\bf r_{gal}}\cdot(-\nabla\Phi(r)_{Z=0}{\tiny })
\end{equation}\\

\begin{figure}[ht!]
 	\begin{center}
 		\includegraphics[width=1.0\textwidth]{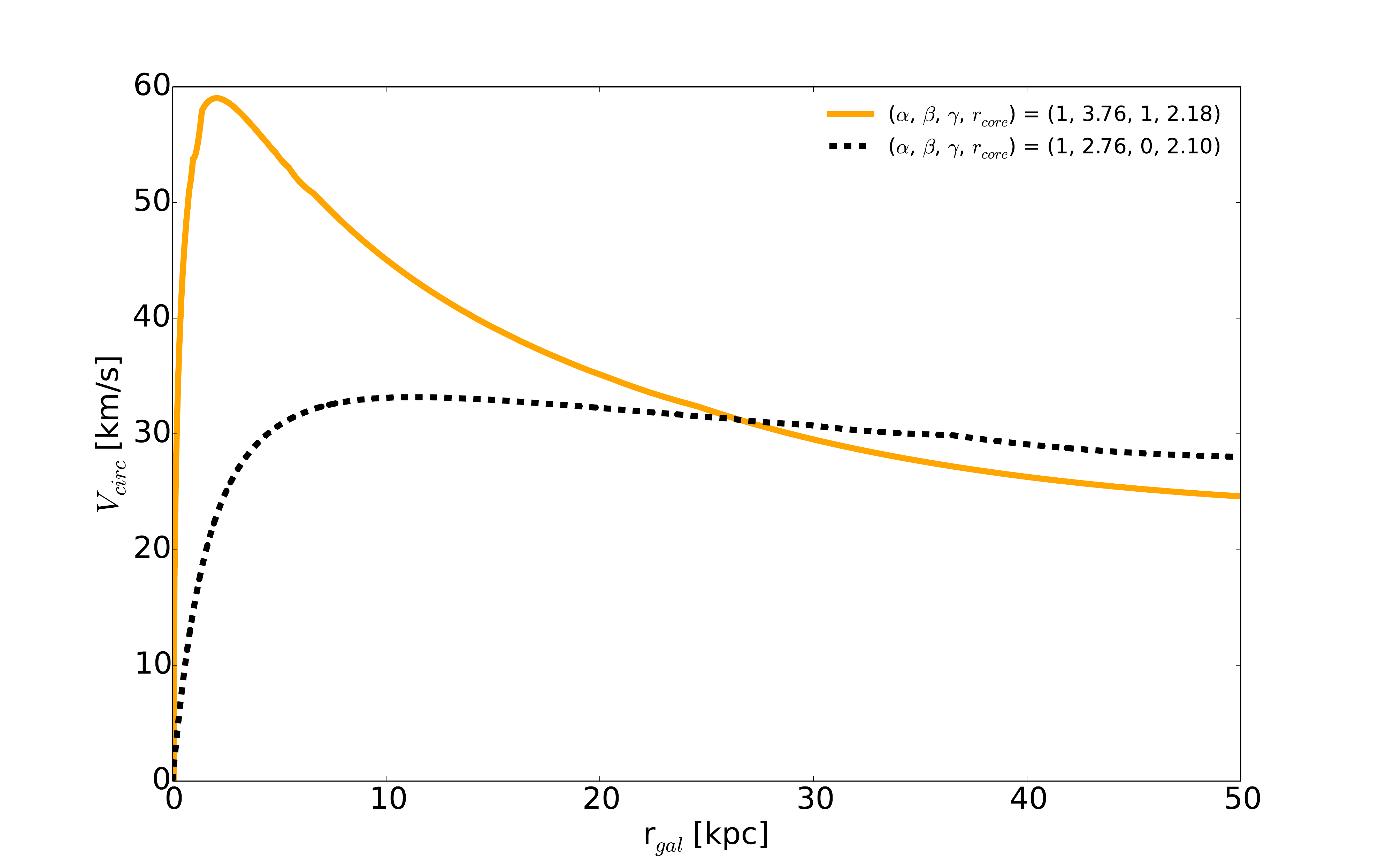}
 \end{center}
 \caption{The contribution to the rotation curve derived from the functional form presented in eq. \ref{generalfunction1}. A double power-law density profile describe 
 	the orange curve, while that a simple power-law density profile reproduces the black dashed line.} 
 	\label{rotationcurve}
 \end{figure}

Finally, the resulting contributions to the rotation curve of the different stellar halo shapes in \S\ref{shapes} are given in Figure \ref{rotationcurve}.\\

  The Besan\c{c}on Galaxy model contains axisymmetric components (including a three-dimensional model for the stellar halo), and a non-axisymmetric structure associated with the triaxial bar. The new dynamical framework of the Galactic model will be presented in forthcoming paper (Fern\'andez-Trincado et al., in preparation).

\begin{acknowledgements}
J.G.F-T is currently supported by Centre National d'Etudes Spatiales (CNES) through Ph.D grant 0101973 and the R\'egion de Franche-Comt\'e, and by the French Programme National de Cosmologie et Galaxies (PNCG). Besan\c{c}on Galaxy Model simulations were executed on computers from the UTINAM Institute of the Universit\'e de Franche-Comte, supported by the R\'egion de Franche-Comt\'e and Institut des Sciences de l'Univers (INSU). Funding for SDSS-III has been provided by the Alfred P. Sloan Foundation, the Participating Institutions, the National Science Foundation, and the U.S. Department of Energy Office of Science. The SDSS-III web site is http://www.sdss3.org/. SDSS-III is managed by the Astrophysical Research Consortium for the Participating Institutions of the SDSS-III Collaboration including the University of Arizona, the Brazilian Participation Group, Brookhaven National Laboratory, Carnegie Mellon University, University of Florida, the French Participation Group, the German Participation Group, Harvard University, the Instituto de Astrofisica de Canarias, the Michigan State/Notre Dame/JINA Participation Group, Johns Hopkins University, Lawrence Berkeley National Laboratory, Max Planck Institute for Astrophysics, Max Planck Institute for Extraterrestrial Physics, New Mexico State University, New York University, Ohio State University, Pennsylvania State University, University of Portsmouth, Princeton University, the Spanish Participation Group, University of Tokyo, University of Utah, Vanderbilt University, University of Virginia, University of Washington, and Yale University.
\end{acknowledgements}

\bibliographystyle{aa}  
\bibliography{Fernandez2} 

\end{document}